\documentclass[aps,preprint,12pt,superscriptaddress,unsortedaddress,floatfix,bibnotes,nofootinbib,unsortedaddress]{revtex4-1}

\usepackage{epsfig}
\usepackage{color}

\newcommand{\eqref}[1]{(\ref{#1})}

\begin{document}

\title{Effective Langevin Equation Approach to the Molecular Diffusion on Optical Lattices}

\author{Aliezer Mart\'{\i}nez-Mesa}
\affiliation{DynAMoS (Dynamical processes in Atomic and Molecular Systems), Facultad de F\'isica, Universidad de la Habana, San L\'azaro y L, La Habana 10400, Cuba}
\author{Llinersy Uranga Pi\~na}
\affiliation{DynAMoS (Dynamical processes in Atomic and Molecular Systems), Facultad de F\'isica, Universidad de la Habana, San L\'azaro y L, La Habana 10400, Cuba}

\begin{abstract}
Optical micro-manipulation techniques has evolved into powerful tools to efficiently steer the motion of microscopical particles on periodic and quasi-periodic potentials, driven by the external electromagnetic field. Here, the dynamics of molecular diffusion on optical lattices is analysed within the framework of the theory of open systems, for polar molecules coupled to a transient electromagnetic field. Using the normal mode expansion of the field, we derive an effective, generalised Langevin equation which describes the motion of the system along the molecular degrees of freedom. The present approach is universally applicable (for molecules with non-vanishing permanent dipole moment) and it opens a wide spectrum of applications in the control of the molecular transport mechanisms on optical lattices. The numerical analysis of suitable model external fields demonstrates the feasibility of neglecting memory terms in the resulting Langevin equation.
\end{abstract}

\maketitle

\section{Introduction}
During the two decades following the experimental realisation of Bose-Einstein condensates (BEC) in atomic gases \cite{Andersson95,Davis95,Bradley95,Fried98,Anglin02}, the confinement, cooling and optical manipulation of atoms and molecules in magnetic and optical traps have attracted a lot of attention. Optical trapping schemes lie at the heart of the continuous progress in the field of cooling and confinement of atoms, and for the collimating of atomic beams. These applications triggered the development of a variety of trapping techniques such as optical tweezers (including beam shaping), optical fibber traps, optical binding, etc. \cite{Paul90,Ketterle96,Petrich95,Grimm00,Cornell91,Anderson95,Ott01,Ovchinnikov97,Weber02,Lee96,Aminoff93,Donley01,Neuman04,Neves15}. 

Furthermore, the low temperatures attainable and the possibility of fine-tuning the parameters determining the shape and depth of the trap \cite{Neves15}, enable the precise monitoring and control of the molecular motion. Likewise, the ability to tune the strength of inter-molecular interactions in a continuous way, via Feshback resonances, provided unparalleled experimental access to a rich body of collective atomic phenomena \cite{Dickerscheid05,Kraemer06}, and it propitiated the advent of quantum simulators \cite{Britton12}.

Optical lattices constitute a periodic generalisation of optical traps. The periodic arrangement of trapping sites is created by counter-propagating laser fields resulting in a standing wave. The three-dimensional periodic pattern resembles the geometry of a crystal, with the additional advantage over real materials of being free of thermal distortions and of structural defects caused by impurities. The key parameters defining the shape of the optical lattices are the well depth, $V_0$, and the spatial periodicity $\lambda$. They can be controlled by modifying the intensity and the wavelength of the electromagnetic field, respectively.

Optical lattices constitute an ideal tool for the investigation of a wide variety of collective phenomena and quantum phase transitions such as Bose-Einstein condensation, the BCS-BEC crossover and the Mott insulator transition \cite{Fabbri09,Clement09,Modugno04,Morsch06,Landig15,Bloch08,Lewenstein12,Struck11}. Moreover, these structures have been widely used in the cooling of atoms and molecules down to nanoKelvin temperatures, the synthesis of new molecules \cite{Neves15}, and they constitute promising candidates for the design and implementation of quantum information and quantum simulation schemes \cite{Jaksch05}. Likewise, the spatial localization of the molecules allows to improve the signal to noise ratio, compared to gas phase spectroscopy, as in matrix isolation spectroscopy experiments. 

One exciting spin-off of the molecular trapping in optical lattices is the possibility to drive their motion through the influence of an additional time-dependent electromagnetic field \cite{Holthaus15,Gemelke05,Sias08,Zenesini09}. The intensity of this time-dependent electromagnetic field is usually much lower than that of the standing wave. This approach have been pursued to control particle diffusion (e.g., to achieve particle localization via the interaction with the field $\vec E(t)$) and to tune the emergence of chaotic behaviour \cite{Steck01,Thommen03,Jones04,Hennequin10,Eckardt17}.
                                                                              
In this Letter, we address the modelling of the classical motion of polar molecules motion on optical lattices in presence of time-dependent electromagnetic fields, and we demonstrate that it can be mapped onto a stochastic process governed by a generalised Langevin equation. To this purpose, we will account for the rapid variations of the electromagnetic field (and their influence on the molecular motion) by using its normal mode decomposition. The electromagnetic field can be regarded as a superposition of plane waves, whose frequencies are typically larger than $1/\tau$ (where $\tau$ is the characteristic time scale of the motion of the molecule on the lattice). From this perspective, molecular diffusion on optical lattices belongs to the group of problems with a clear time scale separation (i.e., comprising both fast and slow degrees of freedom), whose study has a long history in physics, and it constitutes a cornerstone in the system-bath separation in statistical mechanics and thermodynamics \cite{Hasselmann76,Riegert05}.

Although it seems intuitively correct to employ suitable stochastic forces to mimic the influence of transient electromagnetic fields on the molecular diffusion, whether such description remains valid for any waveform is not settled. Up to our knowledge, most studies focused on the control gained over the molecular motion by the manipulation of the coupling between two or more electronic states, induced by a classical or a quantised radiation field \cite{Holthaus15,Gemelke05,Sias08,Zenesini09}. A few investigations addressed the classical description of atomic motion under the influence of a quantised electromagnetic field \cite{Zols97}. Here, we demonstrate that the stochastic model can be derived in the context of classical physics for arbitrary field shapes, and we also provide an analytic formula for the determination of the dissipative memory kernel. Finally, we show that for typical field parameters non-Markovian effects are negligible. Therefore, the numerical analysis of the molecular diffusion can be carried out at a significantly reduced computational cost by using the molecule-field effective friction coefficient.

\section{One-dimensional model}
\label{sec:model}
\subsection{Hamiltonian and equations of motion}
In the following, we consider the motion of a molecule of mass $m$ and permanent electric dipole moment $\vec d$, on a one-dimensional, periodic potential energy curve, e.g., $V(x)=V_0 \cos^2 \left( \lambda x \right)$. The properties of the lasers building up the periodic potential $V(x)$ are not treated explicitly. This information is masked in the controllable parameters $V_0$ and $\lambda$, which allow to take into account the main features of a prototypical optical lattice. The molecule is regarded as a point particle, and the Hamiltonian of the ``system'' is given by
\begin{equation}
 H_S = \frac{p^2}{2m}+V(x). \label{eq:Hs}
\end{equation}
The conclusions derived within this minimal one-dimensional model can be straightforwardly extended to higher dimensional systems.

Within the dipolar approximation, the interaction between the molecule and a time-dependent electromagnetic field $\vec E(x,t)$ is $H_{SB}=-\vec d \cdot \vec E(x,t)$. The field $\vec E(x,t)$ is taken to be a superposition of plane waves, thus the system-bath interaction can be rewritten as
\begin{equation}
 H_{SB} = \sum_j \alpha_j(x) q_j(t) + \sum_j \beta_j(x) \pi_j(t), \label{eq:Hsb}
\end{equation}
where $q_j$ and $\pi_j$ are the generalised coordinate associated to the $j$-normal mode of the electromagnetic field and its conjugate momenta, respectively. The coefficients $\alpha_j(x)$ and $\beta_j(x)$ appearing in equation (\ref{eq:Hsb}) are defined as
\begin{equation}
 \alpha_j(x) = \frac{d\omega_j}{\sqrt{\varepsilon_0 L}}\sin (k_j x), ~~~~~\beta_j(x) = \frac{d}{\sqrt{\varepsilon_0 L}}\cos (k_j x). \label{eq:coef}
\end{equation}
Here, the constants $\varepsilon_0$ and $L$ denote, respectively, the electric permitivity of the vacuum and the quantisation length introduced to define the normal modes of the electromagnetic field.

Likewise, the energy of the field $\vec E(t)$ can be expressed as a superposition of the energies of the normal modes with frequencies $\omega_j$:
\begin{equation}
 H_B = \frac{1}{2} \sum_j \left( \pi_j^2 + \omega_j^2 q_j^2 \right). \label{eq:Hb}
\end{equation}

The total Hamiltonian 
\begin{equation}
 H = \frac{p^2}{2m}+V(x)+\frac{1}{2} \sum_j \left[ \left( \pi_j + \beta_j \right)^2 + \omega_j^2 \left( q_j^2 + \frac{\alpha_j}{\omega_j^2} \right)^2 \right] \label{eq:H}
\end{equation}
can be regarded as a generalisation of the Caldeira-Legget model \cite{Caldeira83}. The main differences with respect to the standard Caldeira-Legget Hamiltonian are the non-linear character of the couplings (mediated by the position-dependent coupling functions $\alpha_j(x)$ and $\beta_j(x)$), and the presence of a term implying a coupling between the coordinate of the particle $x$ and the momenta of the harmonic bath modes.

As in the standard treatment of explicit bath, the Hamiltonian in eq. (\ref{eq:H}) contains the sum $H_S+H_B+H_{SB}$ plus a contribution $\frac{1}{2} \sum_j \left( \beta_j^2 + \alpha_j^2/\omega_j^2 \right)$ that ensures $V(x)$ to be the bare potential along the coordinate $x$.

We now aim to derive an equation that describes the dynamics along the molecular degrees of freedom, upon integration of the normal modes of the transient electromagnetic field. The canonical equations of motion for the molecule can be combined in a single Newton's equation:
\begin{equation}
 m\ddot x + \frac{dV}{dx} = - \sum_j \left( \alpha_j' q_j + \beta_j' \pi_j \right) - \sum_j \left( \beta_j \beta_j' + \frac{\alpha_j \alpha_j'}{\omega_j^2} \right). \label{eq:Newton}
\end{equation}
where $\alpha_j'=d\alpha /dx$ and $\beta_j'=d\beta /dx$.

At the same time, for the bath modes, the Hamilton equations take the form
\begin{eqnarray}
 \dot q_j & = & \pi_j + \beta_j, \nonumber \\
 \dot \pi_j & = & -\omega_j^2q_j-\alpha_j. \label{EOMb}
\end{eqnarray}

The solution of the set of equations (\ref{EOMb}) can be expressed analytically using the Green's function approach:
\begin{equation}
 q_j(t) = Q_{j} (t) + \int_0^t \frac{\left[ \beta_j'(x(t')) \dot x(t') - \alpha_j (x(t')) \right]}{\omega_j} \sin \left[ \omega_j (t-t') \right] dt', \label{eq:qj}
\end{equation}
where
\begin{equation}
 Q_j(t) = q_{j0} \cos \left( \omega_j t \right) +\frac{\pi_{j0}}{\omega_j} \sin \left( \omega_j t \right) \label{eq:capQ}
\end{equation}
describes the harmonic oscillations in absence of system-bath couplings. From eq. (\ref{eq:qj}), the momenta of the bath oscillators can be computed as $\pi_j = \dot q_j - \beta_j$.

\subsection{Langevin equation}
We are interested in the reduced dynamics along the molecular degree of freedom. Inserting the expressions obtained for $q_j(t)$ and $\pi_j(t)$ into equation (\ref{eq:Newton}), we first note, after some algebra, that the terms depending on products of the coupling functions $\alpha_j(x)$, $\beta_j(x)$, and their spatial derivatives $\alpha_j'(x)$, $\beta_j'(x)$, drop out. The resulting equation of motion can be casted in the form of a generalised Langevin equation:
\begin{equation}
 m\ddot x + \frac{dV}{dx} + \xi(x,t) + \frac{\pi |\vec d|^2}{c^2\varepsilon_0 L} \int_0^t F(\tau)\dot x(t-\tau) d\tau = 0. \label{eq:Langevin}
\end{equation}
In this formula, $c$ is the speed of light while the force
\begin{equation}
 \xi(x,t) = \sum_j \left\lbrace \alpha_j'Q_j(t) + \frac{\alpha_j' \alpha_{j0}}{\omega_j^2} \cos (\omega_j t) + \beta_j'\dot Q_j(t)-\frac{\beta_j' \alpha_{j0}}{\omega_j} \sin (\omega_j t ) \right\rbrace \label{eq:xi}
\end{equation}
describes the influence of the ``random'' component of the molecule-field interaction on the molecular motion ($\alpha_{j0}=\alpha_j(x(t=0))$). 

In equation (\ref{eq:Langevin}), $F(\tau)$ is the inverse Fourier transform of the function $\omega^2 \tilde{S}(\omega)$, where $\tilde{S}(\omega)$ is the even continuation of the spectral density of the electromagnetic field:
\begin{equation}
 \tilde{S}(\omega) = 
 \left\{ \begin{array}{ll} 
         |E(\omega)|^2~, \hspace{0.15in} &  \omega \ge 0 \\
         |E(-\omega)|^2~, \hspace{0.15in} &  \omega < 0
         \end{array} \right.
\end{equation}
It can be noticed, that for smooth spectral distributions $|E(\omega)|^2$, the effective bath modes participate in the frictional kernel $F(\tau)$ with the superhomic weight $\sim \omega^2$ characteristic of electromagnetic fields.

Let us note in passing, that conversely to the standard case of dissipative dynamics in mechanical systems, the spectral density $\tilde{S}(\omega)$ contains information on the population of the field normal modes but not on the intensity of the system-bath coupling.

Under the assumption of rapidly decaying kernels, Markov approximation holds, and equation (\ref{eq:Langevin}) can be casted in the form of the standard Langevin equation, for the description of particle-resolved dynamics, or the equivalent Fokker-Planck equation, for the simulation of the time evolution of the probability density in phase space. Within this approximation, the friction coefficient $\gamma$ is given by
\begin{equation}
 \gamma = \frac{\pi d^2}{c^2 \epsilon_0 L} \int_0^{\infty} F(\tau) d\tau.
\end{equation}

\subsection{Approximate and numerical evaluation of the friction kernel}\label{sec:numerical}
To assess the influence of the different parameters determining the shape of the electromagnetic field, on the time-dependence of the friction kernel, we consider a few examples in this section. If the spectral density of the electromagnetic field vanishes outside a vicinity of width $\Delta \omega$ centred a given frequency $\omega_0$, the diffusion coefficient takes the form:
\begin{equation}
F(\tau) = \frac{1}{2} \int_{\omega_0-\Delta\omega/2}^{\omega_0+\Delta\omega/2}\omega^2 |E(\omega)|^2\cos[\omega(t)] d\omega~.
\end{equation}

Moreover, if the function $E(\omega)$ varies smoothly in the interval $\left( \omega_0-\frac{\Delta \omega}{2},\omega_0+\frac{\Delta \omega}{2} \right)$
\begin{equation}
F(\tau) = \omega_0^2 |E(\omega_0)|^2\int_{\omega_0-\Delta\omega/2}^{\omega_0+\Delta\omega/2}\cos[\omega(t)] d\omega = 
\omega_0^2 |E(\omega_0)|^2\cos[\omega_0 \tau]\frac{\sin\left(\frac{\Delta\omega}{2}\tau\right)}{\tau} ~.
\end{equation}
It is straightforward to show, by evaluating the autocorrelation function of the stochastic force $\xi(x,t)$, that the diffusion coefficient in this limit is given by:
\begin{equation}
D = \frac{2|\vec d|^2}{c} \omega_0 |\vec E(\omega_0)|^2\Delta \omega ~.
\end{equation}

\begin{figure}
\centering
\includegraphics[width=0.55\textwidth]{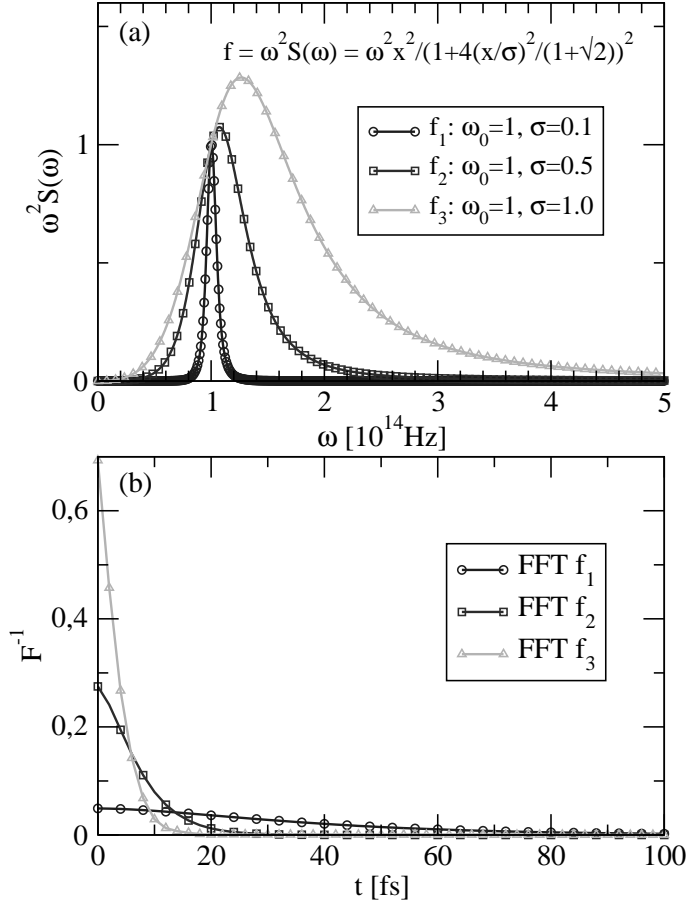}
\caption{(a) Weights $f_i(\omega)$ of the bath oscillators for different widths of a Lorentzian laser pulse with a central frequency of 10$^14$ Hz. (b) Inverse Fast Fourier transforms F$^{−1}$ of the weight functions $f_i(\omega)$.} \label{fig:kernel}
\end{figure}

For radiation fields with a smooth spectral density, the dissipative kernel decays over a period of $\sim 1/\Delta \omega$. If the wavepacket is wide enough in the frequency domain, then this time interval may be much shorter than the characteristic time scale of the molecular motion. We explored numerically the changes in the form of the kernel $F(\tau)$ upon modification of the width and the central frequency of wavepackets of various shapes, in the mid infrared region of the electromagnetic spectrum. In Figure \ref{fig:kernel}, we show the results corresponding to a Lorentzian pulse. They illustrate the general trend of the time evolution of $F(\tau)$ for frequency limited wavepackets, which we also observed in the cases of Gaussian and square-well pulse shapes and for multiperiodic electromagnetic fields.

It can be seen, that for different Lorentzian pulses centred at $\omega_0=10^{14}~$Hz, the friction kernel decays faster as the wavepacket gets broader in spite of the non-standard system-bath couplings in the present case. This behaviour is analogous to that observed in dissipative systems where local-in-time dissipation is obtained for spectral densities of the bath which extend far beyond the characteristic frequency of the system. On the other hand, the characteristic time scale of the friction kernel is independent of the central frequency of the frequency limited pulses. However, the overall magnitude of the dissipative molecule-field coupling significantly enhances as the centre of the wavepacket is shifted to larger frequencies.

\section{Conclusions}
\label{sec:summ}

In summary, we have introduced an stochastic method for studying the molecular diffusion on optical lattices in presence of an external time-dependent electromagnetic field, where the influence of the waveform is filtered into the form of the frictional kernel. The approach is based on the normal mode representation of the electromagnetic field, and the field-molecule interaction is described within the dipole approximation. The method allows to obtain numerically converged results for the diffusion dynamics for arbitrary transient external fields.

The present description is appropriate for wide classes of systems and properties (molecules without a permanent dipole moment constitute nevertheless an important example of systems lying outside the domain of applicability of this methodology). In many situations, the rapid decay of the friction kernel indicates that the numerical integration of the equations of motion may be further simplified by treating the molecular diffusion as a Markov process. In spite of its appeal, to the best of our knowledge, the concept of Langevin dynamics has not been used in the context of laser-driven molecular diffusion on optical lattices.

Although we have focused on an one-dimensional model of the classical motion of molecules on optical lattices, extensions to three-dimensional and quantum systems are straightforward. The present analysis paves the way to employ the mapping of molecular diffusion on optical lattice into a dissipation problem to investigate phenomena such as field-assisted diffusion and tunnelling, taking advantage of the theoretical and computational tools developed over the years to investigate semiclassical (field-free) molecular dynamics. In particular, this methodology can be applied to the control of the diffusive dynamics by tailoring the time-dependent electromagnetic field.

\section*{ACKNOWLEDGEMENTS}

A. M. M. acknowledges the support of the Associate Scheme of the Abdus Salam ICTP.


\begin{thebibliography}{99}
\bibitem{Andersson95} M. H. Andersson, J. R. Ensher, M. R. Matthews, C. E. Wieman, and E. A. Cornell, Science {\bf 269}, 198 (1995)

\bibitem{Davis95} K. Davis, M. O. Mewes, M. R. Andrews, N. J. van Druten, D. S. Durfee, D. M. Kurn, and W. Ketterle, Phys. Rev. Lett. {\bf 75}, 3969 (1995)

\bibitem{Bradley95} C. C. Bradley, C. A. Sacket, J. J. Tollet, and R. G. Hulet, Phys. Rev. Lett. {\bf 75}, 1687 (1995)

\bibitem{Fried98} D. G. Fried, T. C. Killian, L. Willmann, D. Landhuis, S. C. Moss, D. Kleppner, and T. J. Greytak, Phys. Rev. Lett. {\bf 81}, 3811 (1998)

\bibitem{Anglin02} J. R. Anglin, W. Ketterle, Nature \textbf{416}, 211 (2002)

\bibitem{Paul90} W. Paul, Rev. Mod. Phys. {\bf 62}, 531 (1990)

\bibitem{Ketterle96} W. Ketterle and N. J. van Druten, Adv. At. Mol. Opt. Phys. {\bf 37}, 181 (1996)

\bibitem{Petrich95} W. Petrich, M. H. Anderson, J. R. Ensher, and E. A. Cornell, Phys. Rev. Lett. {\bf 74}, 3352 (1995)

\bibitem{Grimm00} R. Grimm, M. Weidem\"uller, and Y. B. Ovchinnikov, Adv. At. Mol. Opt. Phys. {\bf 42}, 95 (2000)

\bibitem{Cornell91} E. Cornell, C. Monroe, and C. Wieman: Phys. Rev. Lett. {\bf 67}, 3049 (1991)

\bibitem{Anderson95} M. H. Anderson, J. R. Ensher, M. R. Matthews, C. E. Wieman, and E. A. Cornell, Science {\bf 269}, 198 (1995)

\bibitem{Ott01} H. Ott, J. Fort\'agh, G. Schlotterbeck, A. Grossmann, and C. Zimmermann, Phys. Rev. Lett. {\bf 89}, 230401 (2001)

\bibitem{Ovchinnikov97} Y. B. Ovchinnikov, I. Manek, and R. Grimm, Phys. Rev. Lett. {\bf 79}, 2225 (1997)

\bibitem{Weber02} T. Weber, J. Herbig, M. Mark, H.-C. Nägerl, and R. Grimm, Science 1079699 (2002)

\bibitem{Lee96} H. J. Lee, C. S. Adams, M. Kasevich, and S. Chu, Phys. Rev. Lett. {\bf 76}, 2658 (1996)

\bibitem{Aminoff93} C. G. Aminoff, A. M. Steane, P. Bouyer, et al., Phys. Rev. Lett. {\bf 71}, 3083 (1993)

\bibitem{Donley01} E. A. Donley, N. R. Claussen, S. L. Cornish, J. L. Roberts, E. A. Cornell, and C. E. Wieman, Nature {\bf 412}, 295 (2001)
\bibitem{Neuman04} K. C. Neuman, S. M. Blocka, Rev. Sci. Instrum. {\bf 75}, 2787 (2004)

\bibitem{Neves15} A. A. R. Neves, P. H. Jones, L. Luo, O. M. Marag\`o, J. Opt. Soc. Am. B \textbf{32}, No. 5 (2015); and references therein

\bibitem{Dickerscheid05} D. B. M. Dickerscheid, U. Al Khawaja, D. van Oosten, H. T. C. Stoof, Phys. Rev. A {\bf 71}, 043604 (2005)

\bibitem{Kraemer06} T. Kraemer, M. Mark, P. Waldburger, J. G. Danzl, et al., Nature {\bf 440}, 315 (2006)

\bibitem{Britton12} J. W. Britton, B. C. Sawyer, A. C. Keith, C. C. J. Wang, J. K. Freericks, H. Uys, M. J. Biercuk, J. J. Bollinger, Nature \textbf{484}, 489 (2012)

\bibitem{Fabbri09} N. Fabbri, D. Cl\'ement, L. Fallani, C. Fort, M. Modugno, K. M. R. van der Stam, and M. Inguscio, Phys. Rev. A {\bf 79}, 043623 (2009)

\bibitem{Clement09} D. Cl\'ement, N. Fabbri, L. Fallani, C. Fort, and M. Inguscio, Phys. Rev. Lett. {\bf 102}, 155301 (2009)

%
\bibitem{Modugno04} M. Modugno, C. Tozzo, and F. Dalfovo, Phys. Rev. A {\bf 70}, 043625 (2004)

\bibitem{Morsch06} O. Morsch, M. Oberthaler, Rev. Mod. Phys. \textbf{78}, 179 (2006)

\bibitem{Landig15} R. Landig, L. Hruby, N. Droga, T. Esslinger, Nature \textbf{532}, 7600 (2015)

\bibitem{Bloch08} I. Bloch, J. Dalibard, W. Zwerger, Rev. Mod. Phys. \textbf{80}, 885 (2008)

\bibitem{Lewenstein12} M. Lewenstein, A. Sanpera, V. Ahofinger, Utracold atoms in optical lattices (Oxford Univ. Press, Oxford, 2012)

\bibitem{Struck11} J. Struck, C. \"Olschl\"ager, R. Le Targat, P. Soltan-Panahi, A. Eckardt, M. Lewenstein, P. Wirdpassinger, K. Sengstock, Science \textbf{333}, 996 (2011)

\bibitem{Jaksch05} D. Jaksch and P. Zoller, Ann. Phys. (N.Y.) {\bf 315}, 52 (2005)

\bibitem{Holthaus15} M. Holthaus, J. Phys. B: At. Mol. Opt. Phys. \textbf{49}, 013001 (2015)

\bibitem{Gemelke05} N. Gemelke, E. Sarajlic, Y. Bidel, S. Hong, S. Chu, Phys. Rev. Lett. \textbf{95}, 170404 (2005)

\bibitem{Sias08} C. Sias, H. Lignier, Y. P. Singh, A. Zenesini, D. Ciampini, O. Morsch, E. Arimondo, Phys. Rev. Lett. \textbf{100} 040404 (2008)

\bibitem{Zenesini09} A. Zenesini, H. Lignier, D. Ciampini, O. Morsch, E. Arimondo, Phys. Rev. Lett. \textbf{102}, 100403 (09)

\bibitem{Eckardt17} A. Eckardt, Rev. Mod. Phys. \textbf{89}, 011004 (2017)

\bibitem{Steck01} D. A. Steck, W. H. Oskay, M. G. Raizen, Science \textbf{293}, 274 (2001)

\bibitem{Thommen03} Q. Thommen, J. C. Garreau, V. Zehnl\'e, Phys. Rev. Lett. {\bf 91}, 210405 (2003)

\bibitem{Jones04} P. H. Jones, M. M. Stocklin, G. Hur, T. S. Monteiro, Phys. Rev. Lett. {\bf 93}, 223002 (2004)

\bibitem{Hennequin10} D. Hennequin, P. Verkerk, Eur. Phys. J. D 57, 95 (2010)

\bibitem{Hasselmann76} K. Hasselmann, Tellus {\bf 28}, 473 (1976)

\bibitem{Riegert05} A. Riegert, N. Baba, K. Gelfert, W. Just, and H. Kantz, Phys. Rev. Lett. {\bf 94}, 054103 (2005)

\bibitem{Zols97} F. Zols, I. Zapata, Effect of QED fluctuations on the dynamics of the macroscopic phase, 403-414, New developments on fundamental problems in quantum physics, M. Ferrero, A. van der Merwe (Kluwer Academic, 1997)

\bibitem{Caldeira83} A. O. Caldeira, A. J. Leggett, Physica A {\bf 121}, 587 (1983)

\end{thebibliography}
\end{document}